\documentstyle[aps,prl,twocolumn,epsfig,graphics,amssymb,epic,eepic]{revtex}

\begin{document}
\onecolumn
\twocolumn[\hsize\textwidth\columnwidth\hsize\csname @twocolumnfalse\endcsname

\title{Interpreting the Wide Scattering of Synchronized Traffic Data by Time Gap Statistics}

\author{Katsuhiro Nishinari$^{1,2,3}$, Martin Treiber$^1$, and Dirk Helbing$^1$} 

\address{$^1$ Institute for Economics and Traffic, Dresden University of
    Technology, 01062 Dresden, Germany\\
$^2$ Department of Applied Mathematics and Informatics, 
Ryukoku University, Shiga 520-2194, Japan\\
$^3$ Institute for Theoretical Physics, University of Cologne, 
50923 K\"oln, Germany
}
\maketitle
\begin{abstract}Based on the statistical evaluation of  experimental single-vehicle data, we
propose a quantitative interpretation of the erratic scattering of flow-density
data in synchronized traffic flows. A correlation analysis suggests that the 
dynamical flow-density data are well compatible with the so-called jam line
characterizing fully developed traffic jams, if one takes into account the variation of 
their propagation speed due to the large variation of the netto time gaps 
(the inhomogeneity of traffic flow). 
The form of the 
time gap distribution depends not only on the density, but also on
the measurement cross section: The most probable netto time gap in congested
traffic flow upstream of a bottleneck is
significantly increased compared to uncongested freeway sections. 
Moreover, we identify different power-law scaling laws for the relative variance of
netto time gaps as a function of the sampling size. While the exponent is
$-1$ in free traffic corresponding to statistically independent time gaps, the
exponent is about $-2/3$ in congested traffic flow because of
correlations between queued vehicles.
\end{abstract}
]
Recently, the statistical physics and non-linear dynamics of driven
many-particle systems have been key disciplines
in the discovery, interpretation, and simulation of phenomena in traffic 
flows \cite{general,Review,TGF01}. The observed instability mechanisms, jamming, segregation,
breakdown, and clustering phenomena are now viewed as a paradigm for similar phenomena
in granular and colloidal physics \cite{TGF01,TGF99,Colloid}, in biology (social insects) \cite{biol},
logistics (instability of supply chains), and economics (business cycles) \cite{econo,TGF99}. 
Empirical investigations \cite{KeRe96,emp,NeubSaScSc99,trap} have particularly 
stimulated the development of traffic theory \cite{trap,theo,Tomer}. 
For example, it was possible to
identify constants of traffic dynamics such as the propagation speed $C \approx -15$~km/h
of wide jams \cite{const1}. According to analytical
calculations by Kerner {\em et al.} for a macroscopic traffic model assuming 
uniformly behaving driver-vehicle units with identical parameters, 
fully developed wide traffic jams should be characterized by a self-organized 
flow-density relation
\begin{equation}
 J(\rho,T,\rho_{\rm max})
 = \frac{1}{T} \left(1 - \frac{\rho}{\rho_{\rm max}} \right) \, ,
\label{form1}
\end{equation}
where $\rho$ denotes the vehicle density, while the average netto time gap
$T$ (the time clearance) and the maximum density $\rho_{\rm max}$ are assumed to be fixed parameters \cite{const2}.  
The dependence of the flow $J$ on the density $\rho$ in Eq. (\ref{form1}) is sometimes referred to as ``jam line'' 
$J(\rho)$ and does not necessarily agree with the high-density part of the steady-state flow-density relation
$Q_{\rm f}(\rho)$ for stationary and homogeneous traffic flow, 
the so-called fundamental diagram \cite{const2}. 
We note, however, that a linear flow-density relation for wide moving jams is not yet fully confirmed by
empirical measurements and that aggregation methods used to determine macroscopic
traffic data from single-vehicle data have a non-trivial impact on the measurement
results \cite{synchro,IDMM}. Nevertheless, the slope $C = dJ(\rho)/d\rho = - 1/(\rho_{\rm max}T)$ 
is usually identified with the average propagation velocity of wide moving jams \cite{const2}.
In contrast to wide moving jams, 
synchronized traffic flow is a form of congested flow which is
mostly found upstream of inhomogeneities 
(e.g. freeway bottlenecks) and claimed to show a completely different behavior \cite{emp}. 
It is characterized by an erratic motion of
time-dependent flow-density data in a two-dimensional area (and a synchronization
of the time-dependent average vehicle
velocities among neighboring lanes) \cite{KeRe96}. More specifically,
in synchronized flow, the slopes 
\begin{equation}
 C(t_{k+1}) = [Q(t_{k+1}) - Q(t_k)]/[\rho(t_{k+1}) - \rho(t_k)] 
\end{equation}
of the lines connecting flow-density data measured at a given freeway cross section
at successive times $t_k$ and $t_{k+1}$ erratically take on positive and negative values, characterizing
a complex spatio-temporal dynamics \cite{KeRe96}.
This is, in fact, one of the most controversial subjects in traffic theory. It
has not only been the reason for Kerner's fundamental criticism of all traffic models
assuming a fundamental (steady-state) relation 
$Q_{\rm f}(\rho)$ between the flow $Q$ and the density $\rho$
(i.e. of the vast majority of models suggested up to now) \cite{critic,general}. 
It has also triggered a flood of publications in physics journals with various suggestions how to 
describe this wide scattering. The proposed interpretations reach from
shock waves propagating forward or backward with speed $C(t)$
\cite{KeRe96,critic}, 
changes in the driving behavior 
in response to the traffic situation (including ``frustration effects'')
\cite{Kra98a,IDMM}, anticipation effects
\cite{Wagn198,KnosSaScSc00}, 
non-unique solutions \cite{Nels100}, a trapping of vehicles \cite{trap},
or multiple metastable oscillating states \cite{Tomer}.
Non-unique solutions are also expected for car-following models, 
in which the vehicle acceleration $dv/dt$ is not a unique function of the
distance or velocity \cite{nonuni}. Other explanations based on 
a mixture of different vehicle types such as cars and
trucks \cite{TreHe99a} or a heterogeneity of
time headways \cite{Bank99} are easier accessible to empirical verification.
In the following, we will therefore focus on these. The particular approach of this study
is its restriction to empirical analyses to avoid theory-driven interpretations. Although
it was been seriously questioned that relation (\ref{form1}) would also be applicable
to flow-density data of synchronized congested traffic flow, by correlation analysis
we will surprisingly find a strong empirical support for the validity of relation (\ref{form1}), 
if the average netto time gap is treated as a dynamical variable.
\par
\begin{figure}[hptb]
\begin{center}
\includegraphics[width=3.7cm, angle=-90]{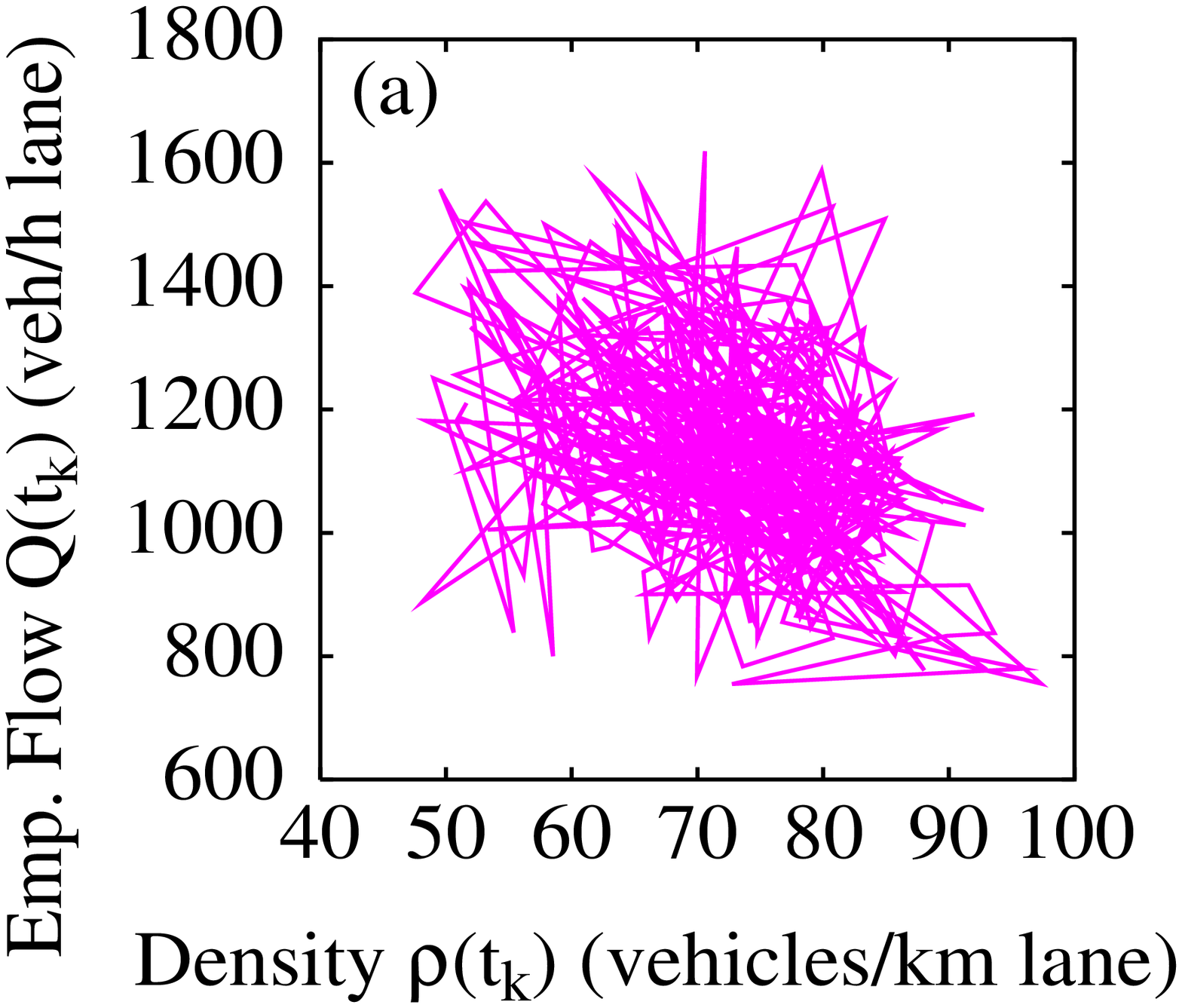}
\includegraphics[width=3.7cm, angle=-90]{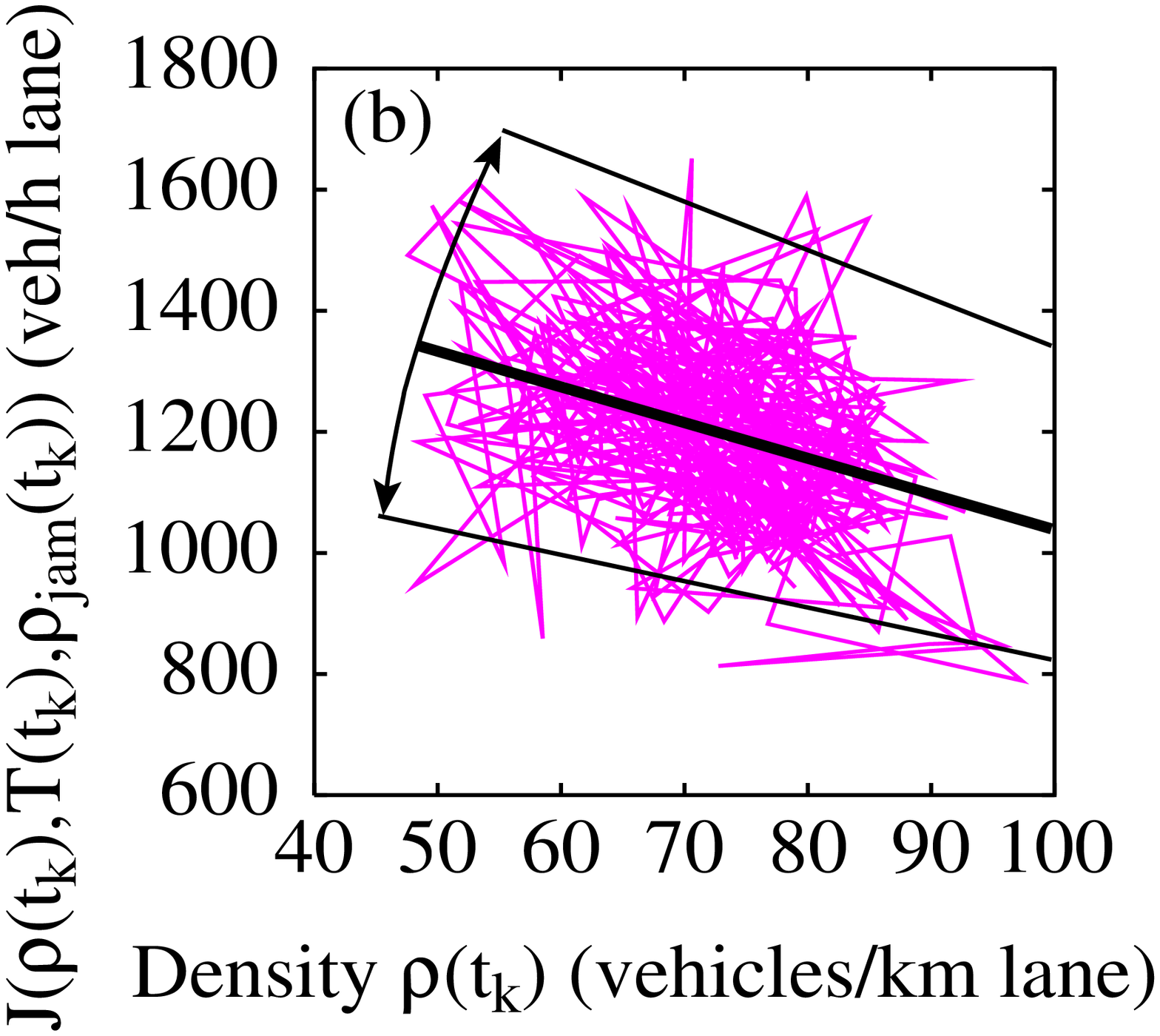}
\end{center}
\caption[]{(Color online)The two-dimensional scattering of empirical
flow-density data in synchronized traffic flow of high density $\rho \ge 45$ veh/km/lane
(see (a)) is well reproduced by the jam relation (\ref{form1}), when not
only the variation of the density $\rho$, but also the empirically measured variation of
the average time gap $T$ and the maximum density $\rho_{\rm max}$ is taken into account (see (b)).
The similarity between the experimental
data and relation (\ref{form1}) is partly because the density $\rho(t_k)$ plotted in 
(a) and (b) (the $x$-value) is determined with the same formula (\ref{seven}), but 
the agreement of the empirical flow $Q = 1/ \langle T_i + l_i/v_i \rangle$ and of the theoretical
relation $J(\rho,T,\rho_{\rm max}) = [ 1-\langle l_i \rangle/\langle v_i (T_i+l_i/v_i)\rangle ]
/\langle T_i \rangle$ (the $y$-values) is not trivial, as even low-order approximations of 
these formulas differ. The pure density-dependence $J(\rho)$ (thick black line) is linear and 
cannot explain a two-dimensional scattering. However, variations of the
average time gap $T$ change its slope $-1/(\rho_{\rm max}T)$ (see arrows), 
and about 95\% of the data are located 
between the thin lines $J(\rho,\overline{T}\pm 2\Delta T,1/l) = (1-\rho l)/(\overline{T}\pm 2 \Delta T)$,
where $l = 3.6$~m is the average vehicle length, $\overline{T}=2.25$~s the average time gap, and
$\Delta T=0.29$~s the standard deviation of $T$.  The predicted form of this area is club-shaped, as
demanded by Kerner \cite{critic}.
\label{Fig1}}
\end{figure}

This requires single-vehicle data, which were recorded by double induction-loop detectors 
for the Dutch freeway A9. The measured data include, for each lane, the
passage time  $t_i^0$ of vehicle $i$, its velocity
$v_i$, and its length $l_i$. From this, we determine the
individual netto time gaps as
\(
T_i = t_i^0 - t_{i-1}^0 - l_i/v_i.
\)
As discussed in Ref.~\cite{TilHe00}, we obtain macroscopic quantities
by averaging over $N=50$ successive vehicles, which gives a better statistics
than averaging over a fixed time interval.
Thus, we obtain the empirical traffic flow 
at time $t_k = \frac{1}{N} \sum_{i=(k-1)N+1}^{kN} t_i^0$
by $Q(t_k)= 1/\langle t_i^0 - t_{i-1}^0 \rangle = N/(t_{kN}^0 - t_{(k-1)N}^0)$
and the average netto time gap by
\begin{equation}
 T(t_k) = \langle T_i \rangle
 = \frac{1}{N} \sum_{i=(k-1)N+1}^{kN} T_i = T_{(k-1)N}^{(N)} \, .
\label{eight}
\end{equation} 
Likewise, we define the vehicle density $\rho$ and the maximum density $\rho_{\rm max}$ by
\begin{equation}
 \frac{1}{\rho(t_k)} 
= \frac{1}{N} \sum_{i=(k-1)N+1}^{kN} v_i (t_i^0 - t_{i-1}^0) \, ,
\label{seven}
\end{equation} 
\begin{equation}
\label{rhojam}
  \frac{1}{\rho_{\rm max}(t_k)} = \langle l_i \rangle = \frac{1}{N} \sum_{i=(k-1)N+1}^{kN} l_i \, .
\end{equation}
For a given loop detector on the left lane, 
we consider all time intervals with
congested traffic (defined by $\rho(t_k) \ge 45$~vehicles per kilometer and lane)
and compare the temporal changes  $\Delta Q^{(k+1)}= [Q(t_{k+1}) - Q(t_k)]$ 
of the empirically measured flow with the changes predicted by Eq. (\ref{form1}).
We will compare three hypotheses by correlation analysis:
(i) $\rho$ is treated as an independent variable defined by (\ref{seven}), while $T$ and $\rho_{\rm max}$ 
are treated as parameters, 
(ii) $\rho$ and $1/T$ are independent variable defined directly from the
single-vehicle data via (\ref{seven}) and (\ref{eight}), while $\rho_{\rm max}$ is a parameter,
(iii) $\rho$, $1/T$, and $1/\rho_{\rm max}$ are independent variables defined by 
(\ref{eight}) to (\ref{rhojam}). The first order Taylor approximations for
the temporal change $\Delta J_{(\alpha)}^{(k+1)}= [J_{(\alpha)}(t_{k+1}) - J_{(\alpha)}(t_k)]$ 
of the corresponding theoretical vehicle flows are obtained
by differential of relation (\ref{form1}) with respect to the independent variables, but not the parameters:
\begin{eqnarray}
\Delta J_{\rm (i)}^{(k+1)}
   &=& - \frac{1}{\rho_{\rm max} T} [\rho(t_{k+1}) - \rho(t_{k})] \, , 
\\
\Delta J_{\rm (ii)}^{(k+1)} 
   &=& \Delta J_{\rm (i)}^{(k+1)}
   + \left[ 1- \frac{\rho(t_k)}{\rho_{\rm max}} \right] 
\left[ \frac{1}{T(t_{k+1})} - \frac{1}{T(t_k)} \right] \, , 
\nonumber \\ 
\Delta J_{\rm (iii)}^{(k+1)} 
   &=& \Delta J_{\rm (ii)}^{(k+1)} - \frac{\rho(t_k)}{T(t_k)}
\left[\frac{1}{\rho_{\rm max}(t_{k+1})}
- \frac{1}{\rho_{\rm max}(t_k)}\right] \, , \nonumber 
\end{eqnarray}
Our statistical analysis
gives the following correlations 
Corr$(\Delta Q,\Delta J_{(\alpha)}) = 
( \sum_{k} \Delta Q^{(k)} \, \Delta J_{(\alpha)}^{(k)})
/[(\sum_{k} \Delta Q^{(k)}{}^2)$
$(\sum_{k} \Delta J_{(\alpha)}^{(k)}{}^2)]^{1/2}$ 
for the hypotheses $\alpha=$i, ii, and iii:
$\mbox{Corr}(\Delta Q,\Delta J_{\rm (i)}) = 0.347$, 
$\mbox{Corr}(\Delta Q,\Delta J_{\rm (ii)}) = 0.918$, 
$\mbox{Corr}(\Delta Q,\Delta J_{\rm (iii)}) = 0.938$. 
As a consequence, hypothesis (i) assuming a linear variation of the flow with the
density is in fact a poor description of ``synchronized'' congested traffic 
data due to their two-dimensional scattering. However, hypothesis (ii)
taking into account the dynamical variation of the average netto time gap $T$ 
yields a strong correlation, which indicates
that we have identified the main reason for the wide scattering
(see Fig.~\ref{Fig1}). Taking into account a variation
of $\rho_{\rm max}$ with the truck fraction improves the correlation coefficient only
a little. Hence, the maximum density, which influences the density-offset of the jam line, 
is an unimportant explanatory variable. 
In contrast, the netto time gap $T$ determining the slope of the jam line (see Eq.~(\ref{form1})) is highly significant.
For this reason, we will investigate the characteristic features of the netto time gap
distribution more closely.
\par
\begin{figure}[hptb]
\begin{center}
\includegraphics[height=4.2cm, angle=-90]{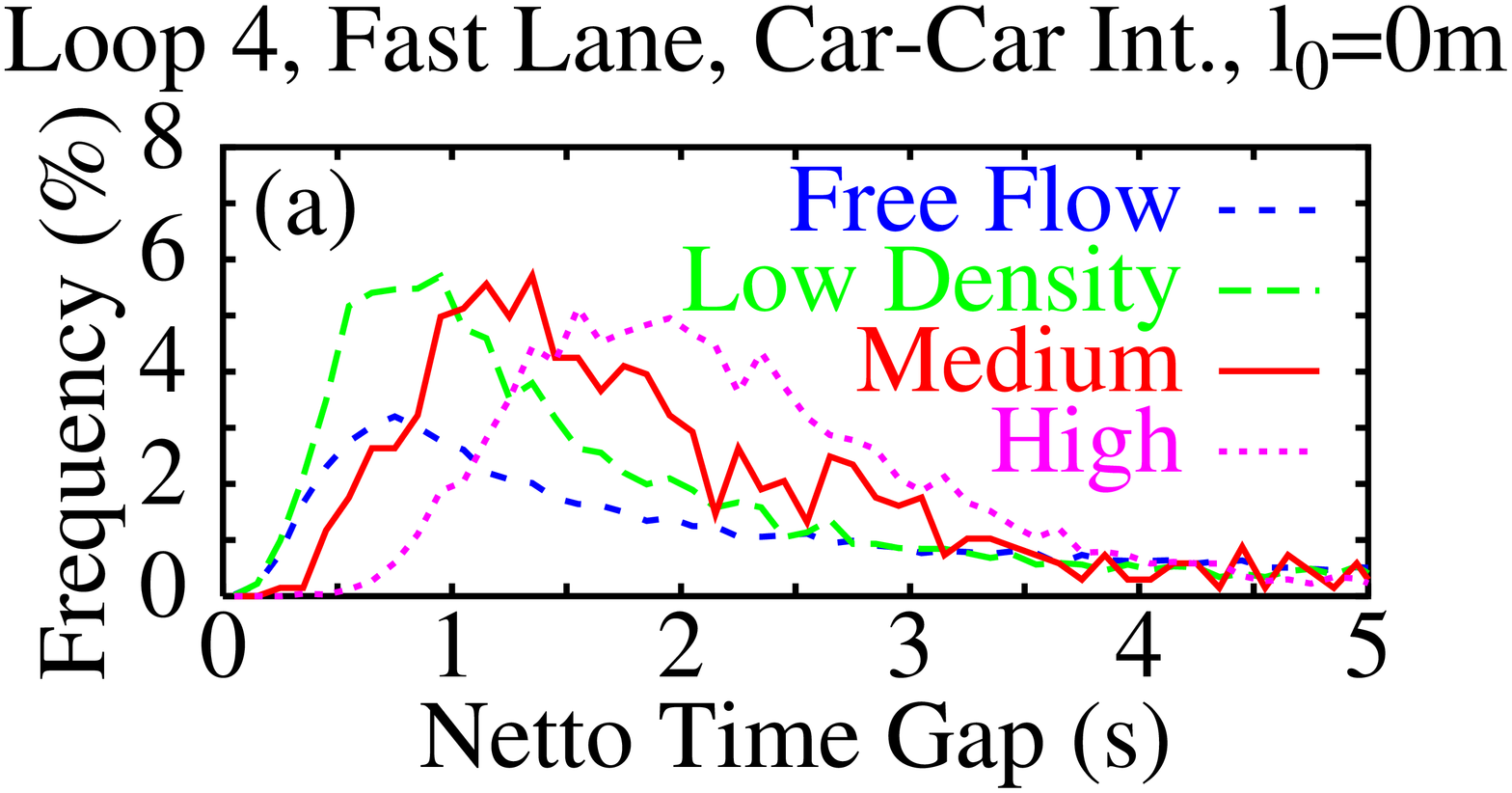}
\includegraphics[height=4.2cm, angle=-90]{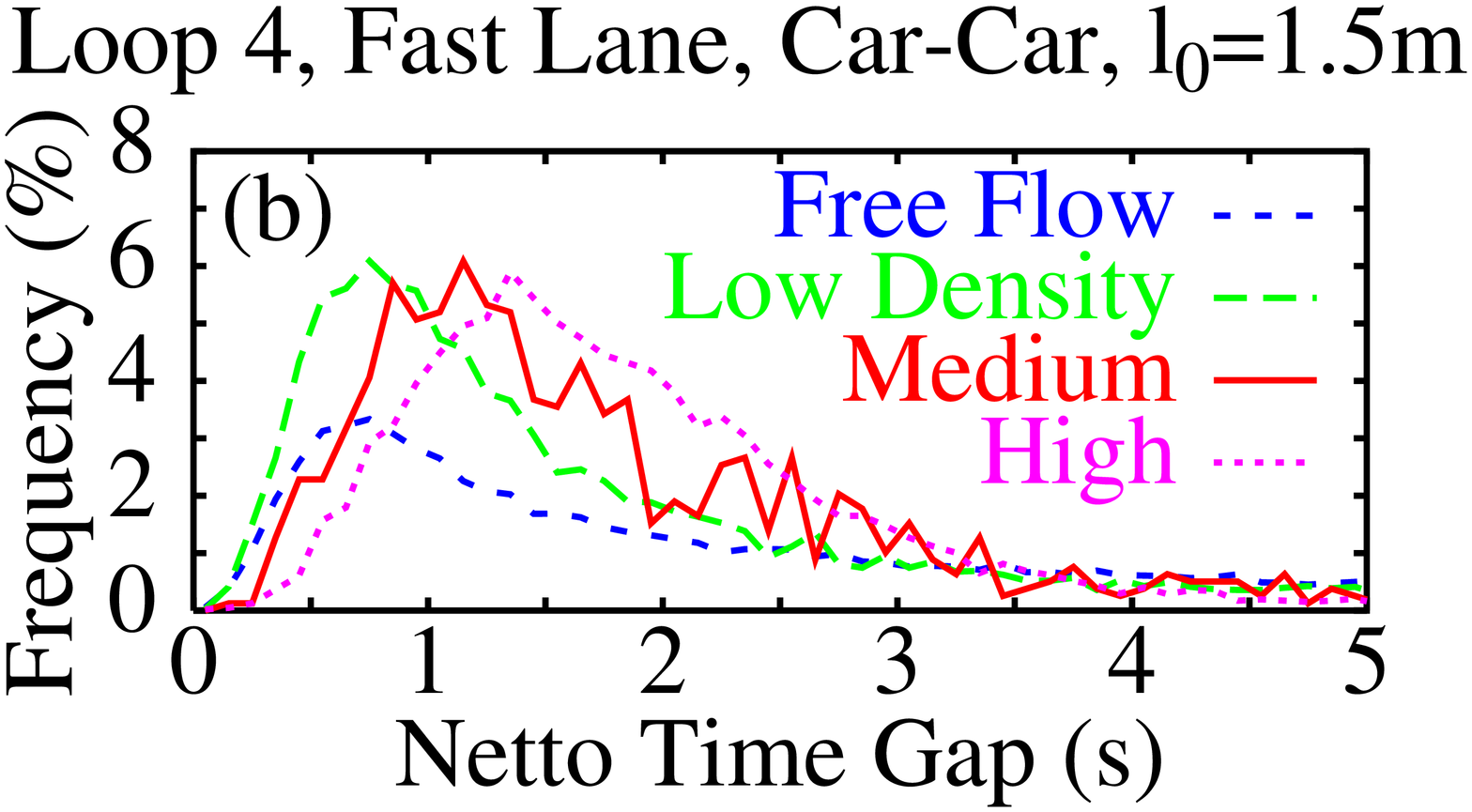}\\
\includegraphics[height=4.2cm, angle=-90]{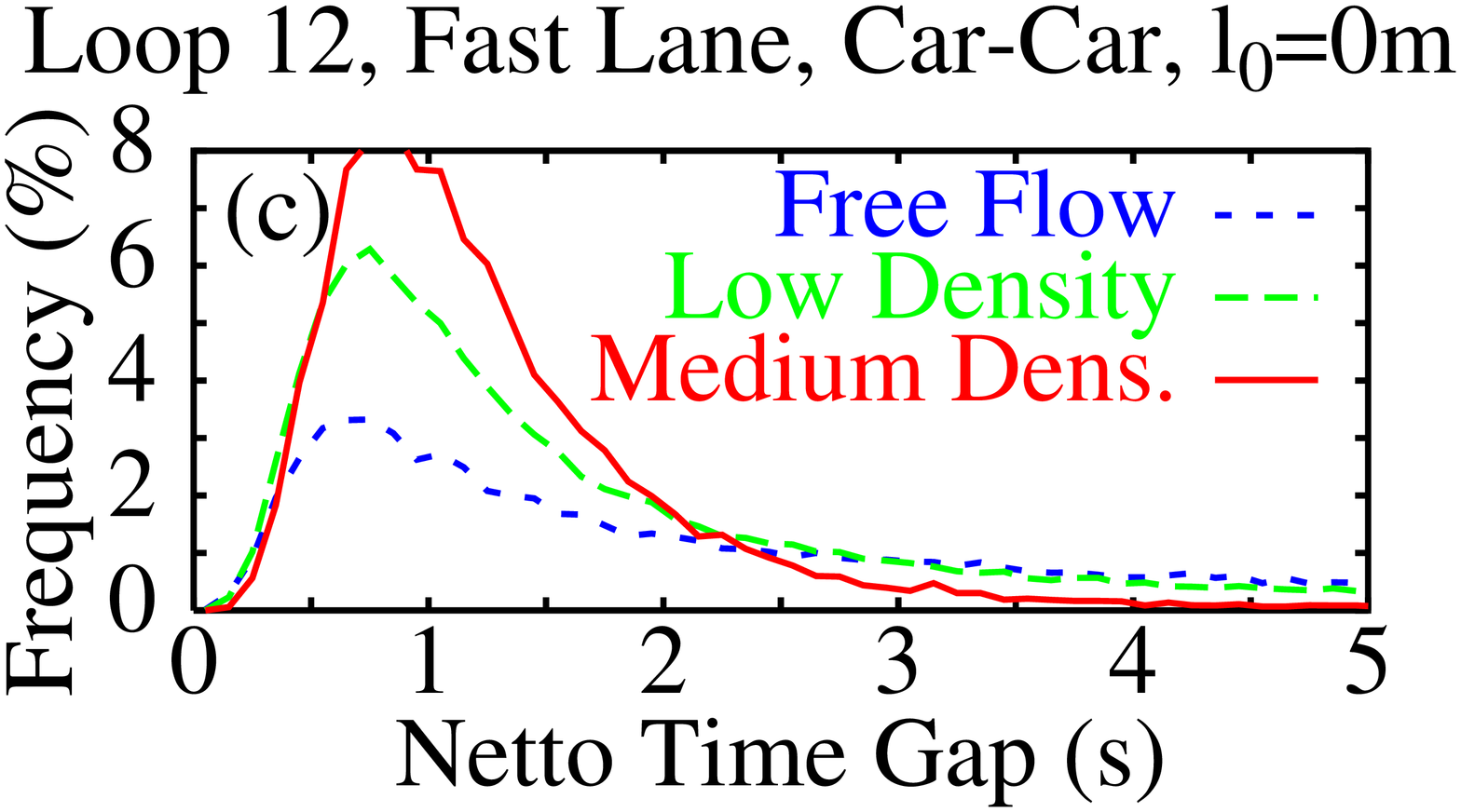}
\includegraphics[height=4.2cm, angle=-90]{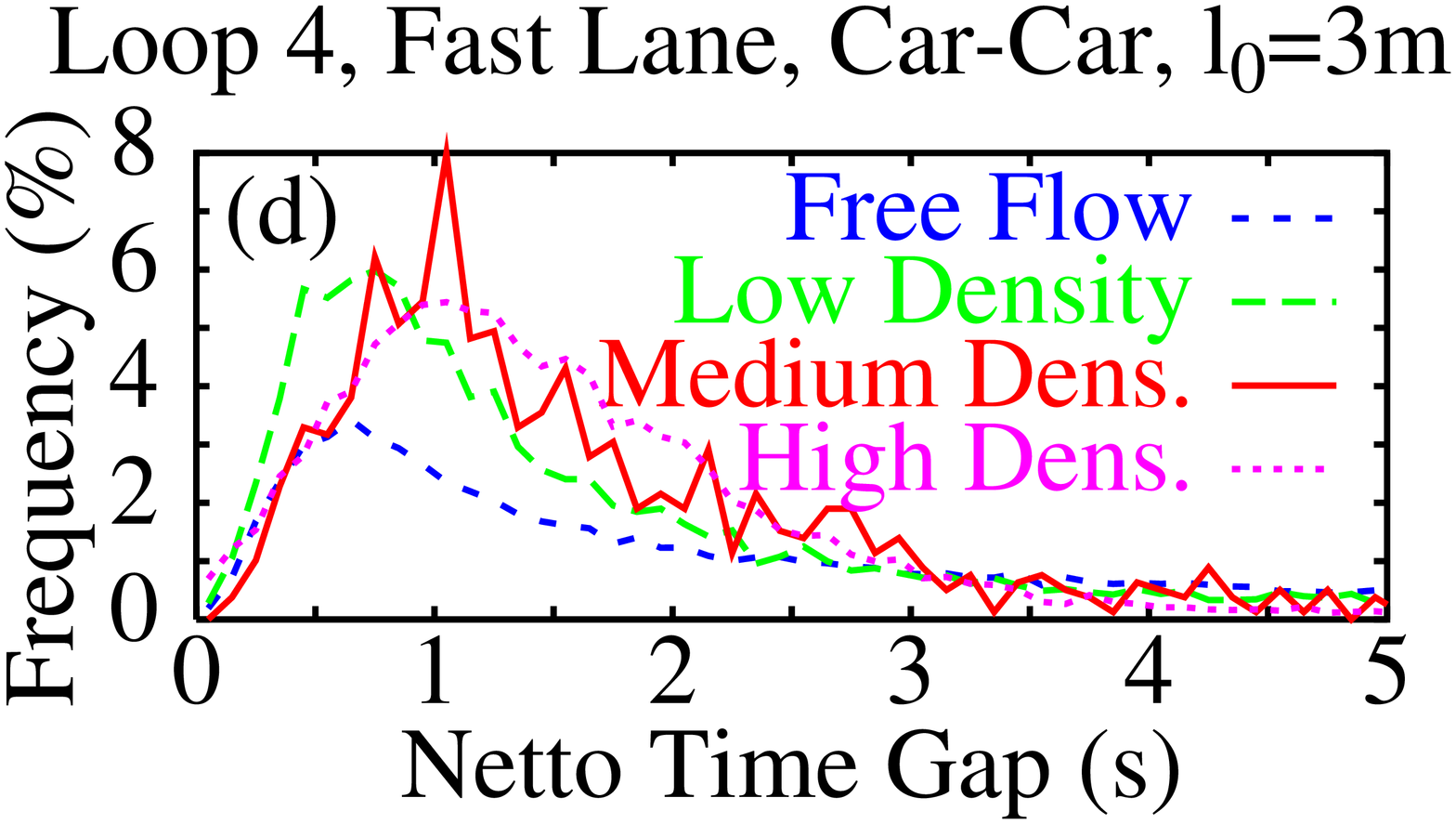}\\
\includegraphics[height=4.2cm, angle=-90]{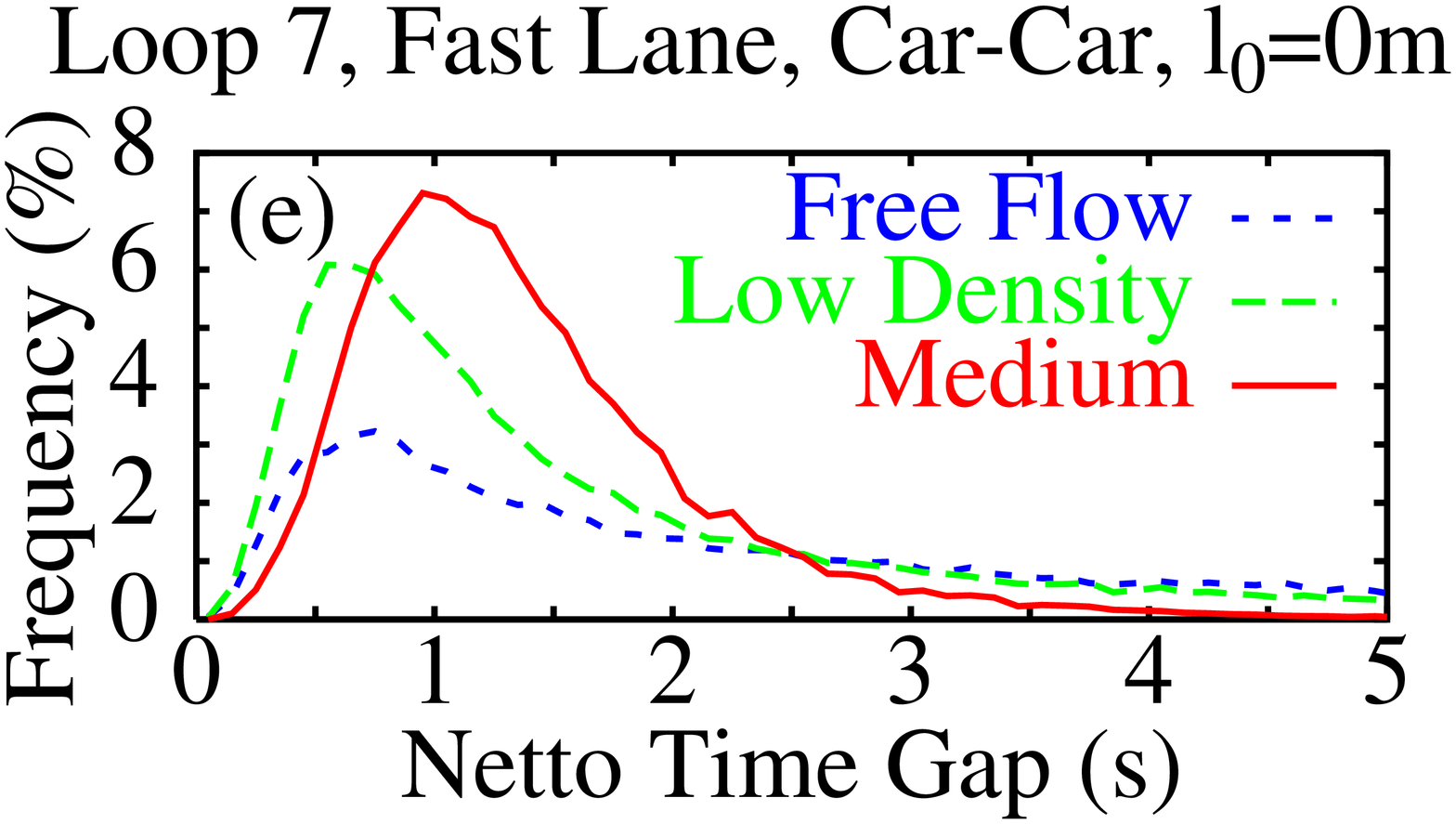}
\includegraphics[height=4.2cm, angle=-90]{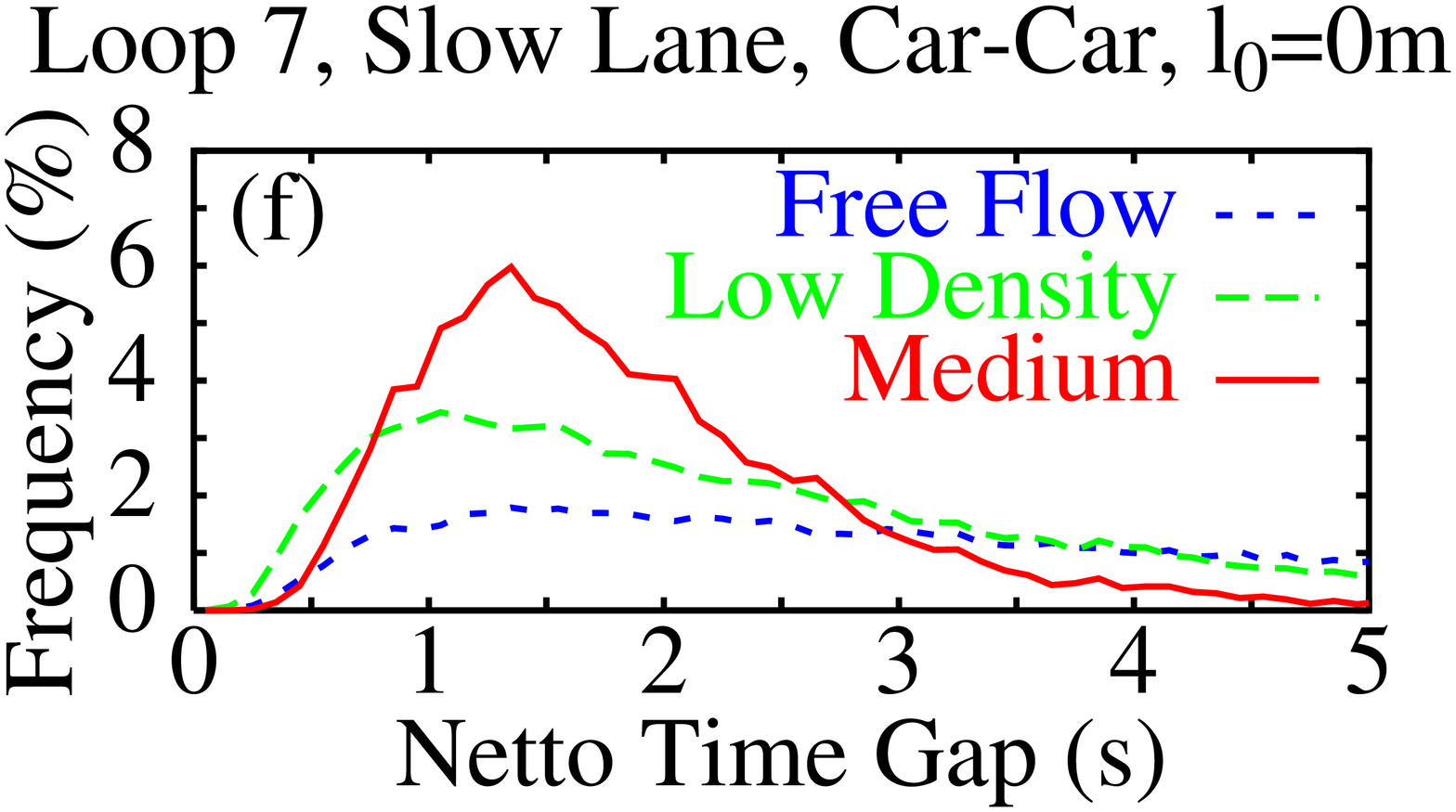}\\
\includegraphics[height=4.2cm, angle=-90]{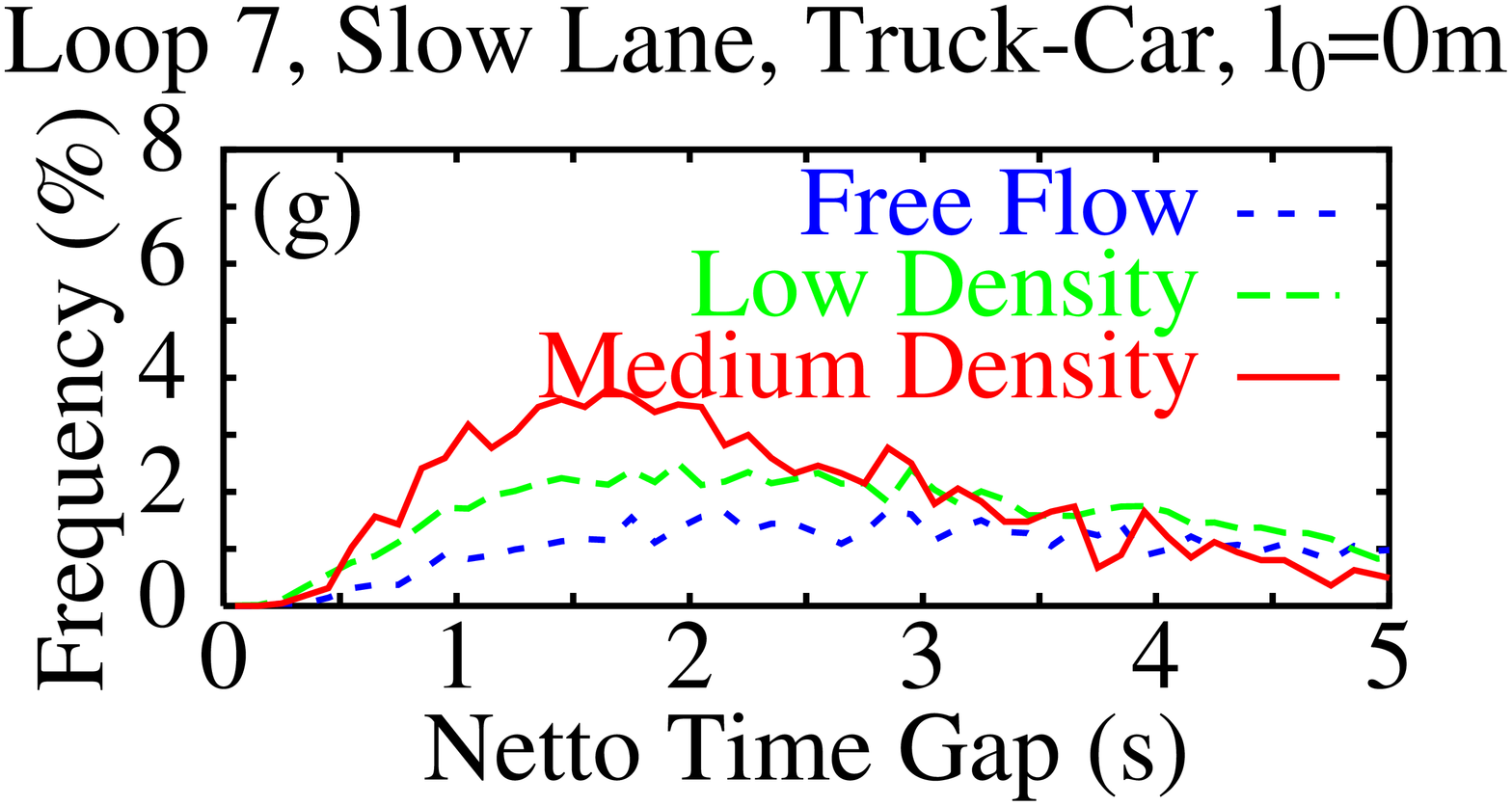}
\includegraphics[height=4.2cm, angle=-90]{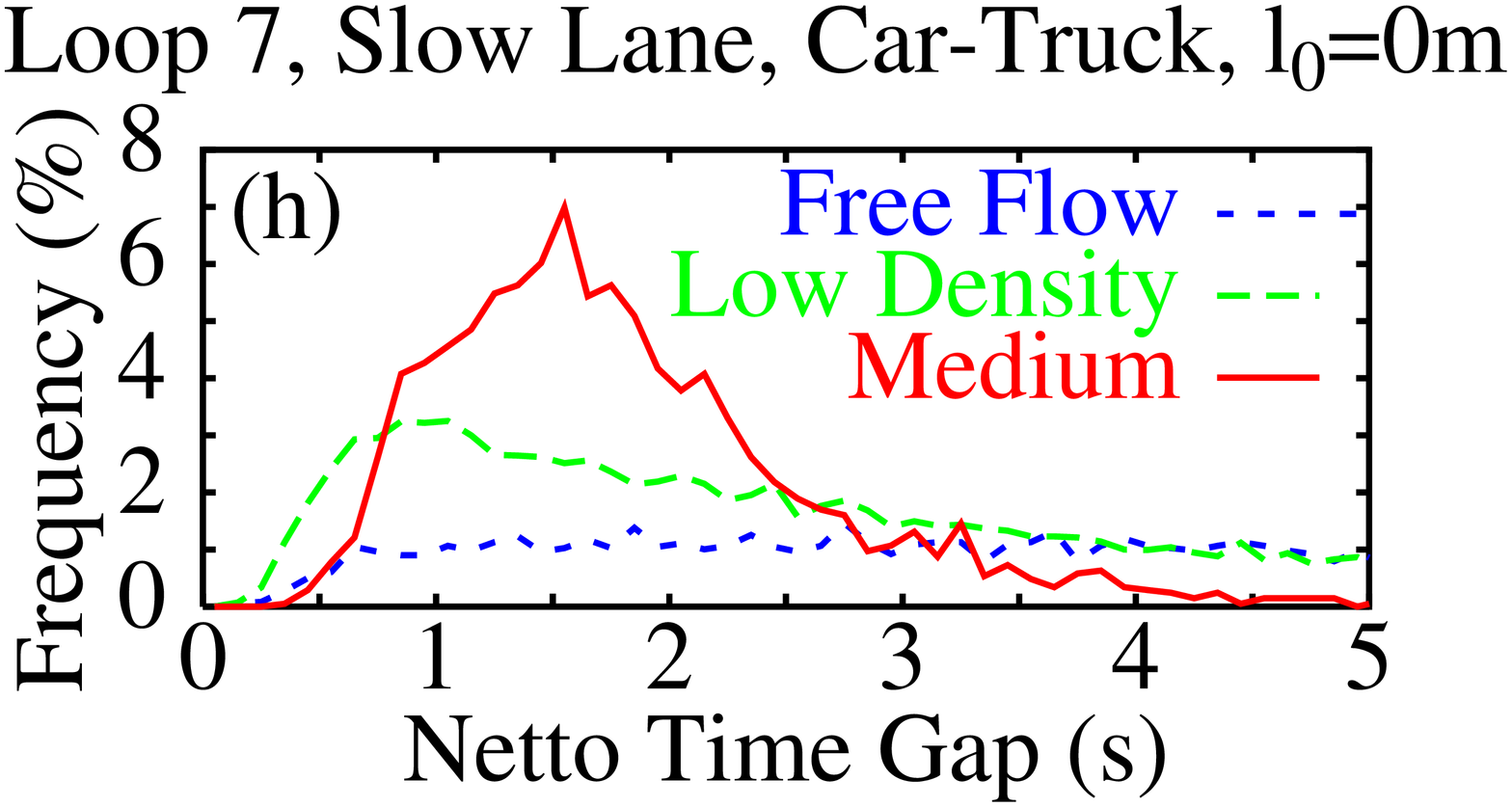}\\
\unitlength0.85cm 
\begin{picture}(9.42,1.8)(-0.1,0.2)
  \thicklines
  \put(4.21,1.2){\vector(1,0){1.0}}
  \thinlines
  \put(0,1.4){\line(1,0){9.42}}

  \put(0,1.2){\line(1,0){0.1}}
  \put(0.2,1.2){\line(1,0){0.1}}
  \put(0.4,1.2){\line(1,0){0.1}}
  \put(0.6,1.2){\line(1,0){0.1}}
  \put(0.8,1.2){\line(1,0){0.1}}
  \put(1.0,1.2){\line(1,0){0.1}}
  \put(1.2,1.2){\line(1,0){0.1}}
  \put(1.4,1.2){\line(1,0){0.1}}
  \put(1.6,1.2){\line(1,0){0.1}}
  \put(1.8,1.2){\line(1,0){0.1}}
  \put(2.0,1.2){\line(1,0){0.1}}
  \put(2.2,1.2){\line(1,0){0.1}}
  \put(2.4,1.2){\line(1,0){0.1}}
  \put(2.6,1.2){\line(1,0){0.1}}
  \put(2.8,1.2){\line(1,0){0.1}}
  \put(3.0,1.2){\line(1,0){0.1}}
  \put(3.2,1.2){\line(1,0){0.1}}
  \put(3.4,1.2){\line(1,0){0.1}}
  \put(3.6,1.2){\line(1,0){0.1}}
  \put(3.8,1.2){\line(1,0){0.1}}
  \put(4.0,1.2){\line(1,0){0.1}}
  \put(5.3,1.2){\line(1,0){0.1}}
  \put(5.5,1.2){\line(1,0){0.1}}
  \put(5.7,1.2){\line(1,0){0.1}}
  \put(5.9,1.2){\line(1,0){0.1}}
  \put(6.1,1.2){\line(1,0){0.1}}
  \put(6.3,1.2){\line(1,0){0.1}}
  \put(6.5,1.2){\line(1,0){0.1}}
  \put(6.7,1.2){\line(1,0){0.1}}
  \put(6.9,1.2){\line(1,0){0.1}}
  \put(7.1,1.2){\line(1,0){0.1}}
  \put(7.3,1.2){\line(1,0){0.1}}
  \put(7.5,1.2){\line(1,0){0.1}}
  \put(7.7,1.2){\line(1,0){0.1}}
  \put(7.9,1.2){\line(1,0){0.1}}
  \put(8.1,1.2){\line(1,0){0.1}}
  \put(8.3,1.2){\line(1,0){0.1}}
  \put(8.5,1.2){\line(1,0){0.1}}
  \put(8.7,1.2){\line(1,0){0.1}}
  \put(8.9,1.2){\line(1,0){0.1}}
  \put(9.1,1.2){\line(1,0){0.1}}
  \put(9.3,1.2){\line(1,0){0.1}}

  \put(1,0.5){\line(0,1){0.9}}
  \put(2.06,0.5){\line(0,1){0.9}}
  \put(2.56,0.5){\line(0,1){0.9}}
  \put(3.01,0.5){\line(0,1){0.5}}
  \put(3.51,0.5){\line(0,1){0.9}}
  \put(4.71,0.5){\line(0,1){0.9}}
  \put(6.71,0.5){\line(0,1){0.9}}
  \put(7.41,0.5){\line(0,1){0.9}}
  \put(7.71,0.5){\line(0,1){0.9}}
  \put(8.42,0.5){\line(0,1){0.9}}
  \put(0.1,0.7){\makebox(0,0){\footnotesize (i)}}
  \put(0.25,0.2){\makebox(0,0){\footnotesize $x=$}}
  \put(1.00,0.2){\makebox(0,0){\footnotesize 43.3}}
  \put(1.75,0.2){\makebox(0,0){\footnotesize 42.3}}
  \put(2.45,0.2){\makebox(0,0){\footnotesize 41.8}}
  \put(3.12,0.2){\makebox(0,0){\footnotesize 41.3}}
  \put(3.85,0.2){\makebox(0,0){\footnotesize 40.8}}
  \put(4.71,0.2){\makebox(0,0){\footnotesize 39.6}}
  \put(6.44,0.2){\makebox(0,0){\footnotesize 37.6}}
  \put(7.24,0.2){\makebox(0,0){\footnotesize 36.9}}
  \put(7.94,0.2){\makebox(0,0){\footnotesize 36.6}}
  \put(9.0,0.21){\makebox(0,0){\footnotesize 36.9\ km}}
 
  \put(0.36,1.7){\makebox(0,0){\footnotesize Loop}}
  \put(1.00,1.7){\makebox(0,0){\footnotesize 3}}
  \put(2.05,1.7){\makebox(0,0){\footnotesize 4}}
  \put(2.55,1.7){\makebox(0,0){\footnotesize 5}}
  \put(3.01,1.7){\makebox(0,0){\footnotesize 6}}
  \put(3.58,1.7){\makebox(0,0){\footnotesize 7}}
  \put(4.71,1.7){\makebox(0,0){\footnotesize 8}}
  \put(6.70,1.7){\makebox(0,0){\footnotesize 9}}
  \put(7.29,1.7){\makebox(0,0){\footnotesize 10}}
  \put(7.79,1.7){\makebox(0,0){\footnotesize 11}}
  \put(8.40,1.7){\makebox(0,0){\footnotesize 12}}
  \put(0,1){\line(1,0){0.75}}
  \put(0.75,1){\line(1,-1){0.15}}
  \put(1,1){\line(1,-1){0.15}}
  \put(1,1){\line(1,0){1.06}}
  \put(2.06,1){\line(-1,-1){0.15}}
  \put(2.31,1){\line(-1,-1){0.15}}
  \put(2.31,1){\line(1,-1){0.15}}
  \put(2.56,1){\line(1,-1){0.15}}
  \put(2.56,1){\line(1,0){0.45}}
  \put(3.01,1){\line(-1,-1){0.15}}
  \put(3.26,1){\line(-1,-1){0.15}}
  \put(3.26,1){\line(1,0){3.9}}
  \put(7.16,1){\line(1,-1){0.15}}
  \put(7.41,1){\line(1,-1){0.15}}
  \put(7.41,1){\line(1,0){0.3}}
  \put(7.71,1){\line(-1,-1){0.15}}
  \put(7.96,1){\line(-1,-1){0.15}}
  \put(7.96,1){\line(1,0){1.46}}
\end{picture}

\end{center}
\caption[]{(Color online)(Netto) time gap distributions at different cross sections
of the Dutch freeway A9 (see the sketch in subfigure (i)), separately measured 
for the fast and slow lane, for car-car,
car-truck, and truck-car interactions, and 
for different density regimes, namely
free flow ($\rho < 10$ veh/km/lane), low densities (free traffic with 
$10 < \rho < 25$ veh/km/lane),
medium densities (mostly congested traffic with $25 < \rho < 50$ veh/km/lane), and high densities
(congested traffic with $\rho > 50$ veh/km/lane, only observed upstream of serious bottlenecks). 
Densities were determined as 50-vehicle averages, and 
trucks were defined by a vehicle length $l_i\ge 6$ m.  The time gaps of trucks with respect
to cars are, on average, considerably higher than the ones of cars with respect to other vehicles,
which causes a large variation of individual time gaps. However, even the variation of time
gaps among cars is considerable. 
A remarkable point is 
the increase in the most probable time gap from 0.75~s in free traffic 
(on the fast lane) to considerably higher values in congested traffic
upstream of a bottleneck (1.2~s at medium densitities and 1.9~s at high densities
at Loop 4 compared to 0.75~s at Loop 12 in all density regimes and 1~s at Loop 7 at
medium densities, see (a), (c), (e); the values for the slow lane are somewhat higher).
This ``frustration effect'' is still slightly
active immediately downstream of a bottleneck (Loop 7), in contrast to freeway sections,
which are never seriously congested (Loop 12). It is also observed for
finite minimum bumper-to-bumper distances of, for example,
$l_0 = 1.5$ or 3~meters, which is considered by replacing $l_i$ by  $(l_i+l_0)$ everywhere.
\label{Fig2}}
\end{figure}
\begin{figure}[hptb]
\begin{center}
\includegraphics[width=4cm, angle=-90]{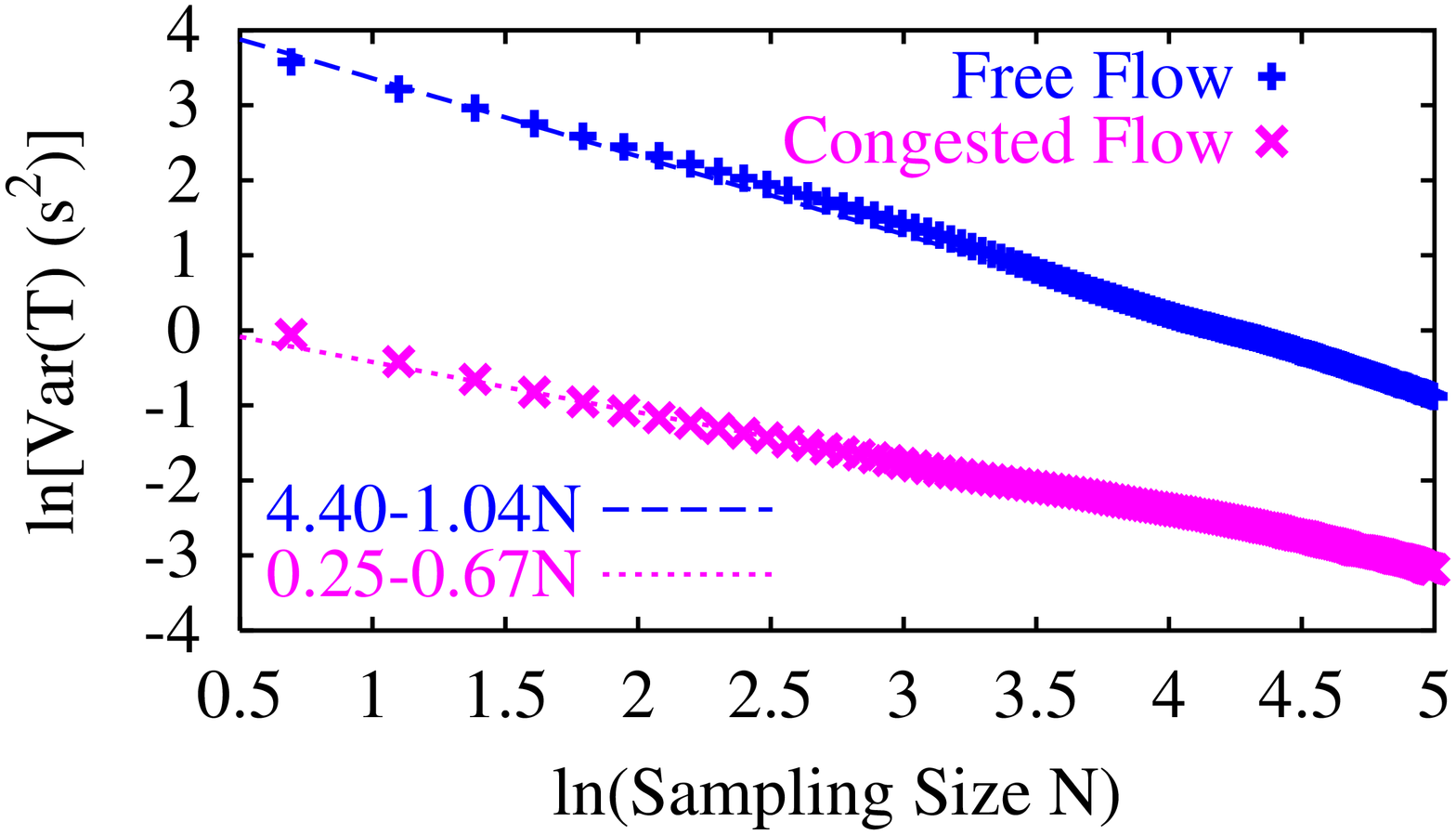}
\end{center}
\caption[]{(Color online)Variance of the average time gaps $T$ as a function of the sampling size $N$,
separately measured for free and congested flows at loop 4 on the fast lane. While in free flow,
the relative variance is consistent with the expected power law exponent of $\gamma = -1$,
it decreases much slower with $N$ in the 
congested traffic regime, namely with an exponent of $\gamma \approx -0.67$,
causing the significant variation of $T$.\label{Fig3}}
\end{figure}

The variation of the average time gaps with time can only be relevant, when it is large.
The time gap distribution is, in fact, surprisingly wide, and it has so-called heavy tails
(see Fig.~\ref{Fig2}). 
To investiagte this, let us calculate the variance of the sample-averaged
time gaps as a function of the sampling size $N$,
\begin{equation}
\label{varT}
 \mbox{Var}(T) = \frac{1}{I-N+1} \sum_{j=0}^{I-N}
 \left(T_{j}^{(N)} - \bar{T}\right)^2 \, ,
\end{equation}
where $(j+1)$ runs over the first vehicle indices of all possible
samples of $N$ consecutive vehicles in the data set of size $I$,
$T_{j}^{(N)}=\frac{1}{N}\sum_{i=j+1}^{j+N} T_j$ 
is defined as moving average of the time gaps of the 
$N$ next vehicles,
and $\bar{T}=\sum_{i=1}^I T_i/I = T_0^{(I)}$ is the global average.
To account for artifacts caused by the daily variation of traffic flow,
we applied a high-pass filter (with cutoff period
$N_c=500$) to the single-vehicle data before
calculating Eq. (\ref{varT}), which  limits the scaling range
of relation (\ref{varTN}). 
For a given  $N$, the variance (\ref{varT}) decreases with increasing density 
(see Figs.~\ref{Fig2} and \ref{Fig3}),
as less and less space is available for time gaps larger than the 
safe time gap, but 
the variance is also a function of the sampling size $N$. 
For free traffic flow (with $\rho < 15$ veh/km/lane),  
we observe the power law behavior $\mbox{Var}(T) \propto 1/N$, 
as expected for
statistically independent time gaps in free traffic. At high vehicle densities
(with $\rho > 35$ veh/km/lane), however, we find the power law 
\begin{equation}
\label{varTN}
 \mbox{Var}(T) \propto N^{\gamma}
\end{equation}
with an exponent $\gamma \approx -0.67$ (see Fig.~\ref{Fig3}). 
That is, the relative variance decreases much slower with the sampling
size than expected, implying that {\em the time gaps do not average well and fluctuations
of average time gaps remain significant for reasonable sampling sizes $N$.}  
This is related with surprisingly wide time gap distributions
and results from correlations between queued vehicles, which are probably due to 
platoon formation \cite{trap}, but may also be caused by dynamical 
vehicle interactions. The reproduction of this scaling law will be a serious test for 
microscopic traffic simulation models, which are needed to reveal the detailed mechanism behind it.
\par
According to Figure~\ref{Fig2}, the time gap distribution is not only 
dependent on the density, but also on the measurement cross section (more specifically:
on the relative location with respect to bottlenecks of the freeway). It has been 
suggested \cite{Bank91} that the increase of the most probable {\em brutto} time gap 
(time headway) $\Delta t_i^0 = t_i^0 - t_{i-1}^0 = T_i + l_i/v_i$ in congested traffic compared to 
free traffic \cite{NeubSaScSc99,TilHe00} 
is due to the drop of the vehicle velocity $v_i$, increasing $l_i/v_i$. However, an increase
is also observed for the most probable {\em netto} time gaps $T_i$ 
(see Fig.~\ref{Fig2}). This increase is
most pronounced at cross sections upstream of bottlenecks (e.g. on-ramps), 
where vehicles are queueing. 
We interpret this 
increase of the netto time gaps as {\em ``frustration effect''} of drivers after a considerable 
queuing time, which is supported by other observations as well
\cite{TGF01,gapdis}. The frustration effect is most significant at freeway sections, where the
accumulated time of driving under congested conditions is high, and it decays with
increasing recovery time. At freeway sections, where serious congestion never occurs,
there is no relevant increase in the most probable netto time gap with growing density
(see Fig.~\ref{Fig2}). 
Finally, we note that the time gaps in front of long vehicles (``trucks'') are much higher, on
average, than in front of short vehicles (``cars''), as expected (see Fig.~\ref{Fig2}). 
This supports the theory
suggested in Ref.~\cite{TreHe99a}. 
\par
In conclusion, we have found an interpretation of the wide scattering of flow-density 
data in ``synchronized'' congested traffic based on the jam relation $J(\rho)$, but
taking into account the time-dependent variation
of empirical netto time gaps among successive cars. This variation is related to time-dependent 
changes of the slope $C=-1/(\rho_{\rm max}T)$ of the jam line and causes the two-dimensional scattering
together with the variation of the vehicle density. 
The surprisingly strong variation of the average
time gaps was due to the fact that 
the power law relating the variance of the time
gaps with the sampling size had a considerably smaller exponent ($\gamma \approx -2/3$)
than expected ($\gamma = -1$). Hence, distinguishing different forms of congested
traffic based on the scattering is questionable. We also found that the increase in
the most probable time gap in congested traffic is site-dependent and not attributed to
a drop in the vehicle velocity, indicating a ``frustration effect''. Traffic models
considering the dynamical changes of the netto time gaps can reproduce empirical
observations more realistically \cite{TreHe99a,IDMM}. 

{\em Acknowledgements:} The authors would like to thank
Henk Taale and the {\em Dutch Ministry of Transport, Public Works
and Water Management} for providing the 
induction double-loop single-vehicle detector data.\\[-8mm]

\end{document}